\documentclass[sigconf,nonacm]{acmart}
\AtBeginDocument{%
  }

\setcopyright{acmlicensed}
\copyrightyear{2024}
\acmYear{2024}
\acmDOI{XXXXXXX.XXXXXXX}

\acmConference[AIOps '24]{the 5th International Workshop on Cloud Intelligence and AIOps Co-located with ASPLOS 2024}{April 27,
  2024}{San Diego, CA}

\acmISBN{978-1-4503-XXXX-X/18/06}

\usepackage{tikz}     % to draw some self-contained figs
\usepackage{amsmath}
\usepackage{xspace}
\newcommand{\xxx}{SSJF\xspace}

\usepackage{multirow, multicol, booktabs, tabulary, tabu, longtable, array, varwidth}
\usepackage{placeins, lipsum}
\usepackage[flushleft]{threeparttable}
\usepackage{tablefootnote}

\usepackage{tikz}
\usepackage{xcolor}

\usepackage[]{cleveref} % get fancy referencing
\crefname{appsec}{Appendix}{Appendices}
\crefformat{appendix}{Appendix~#2#1#3}
\crefname{definition}{Def.}{Defs.}
\crefformat{section}{\S#2#1#3}
\crefrangeformat{section}{\S#3#1#4--\S#5#2#6}
\crefformat{subsection}{\S#2#1#3}
\crefformat{subsubsection}{\S#2#1#3}
\crefrangeformat{subsection}{\S#3#1#4--\S#5#2#6}
\crefmultiformat{subsection}{\S#2#1#3}{ and~\S#2#1#3}{, \S#2#1#3}{ and~\S#2#1#3}
\crefformat{equation}{Eq. #2#1#3}
\crefrangeformat{equation}{(#3#1#4--#5#2#6)}
\crefmultiformat{equation}{(#2#1#3)}{ and~(#2#1#3)}{, (#2#1#3)}{ and~(#2#1#3)}
\crefformat{figure}{Fig.~#2#1#3}
\crefrangeformat{figure}{Figs. #3#1#4--#5#2#6}
\crefmultiformat{figure}{Figs.~#2#1#3}{ and~#2#1#3}{, #2#1#3}{ and~#2#1#3}
\crefformat{algorithm}{Alg.~#2#1#3}
\crefformat{table}{Table~#2#1#3}
\crefrangeformat{table}{Tables~#3#1#4--#5#2#6}
\crefmultiformat{table}{Tables~#2#1#3}{ and~#2#1#3}{, #2#1#3}{ and~#2#1#3}

\usepackage{enumerate}

\usepackage[inline]{enumitem}
\setlist{noitemsep,nolistsep,leftmargin=*}

\usepackage{algorithm}
\usepackage{algpseudocode}

\usepackage[compact]{titlesec}
\titlespacing*{\section}{0pt}{6pt plus 4pt minus 2pt}{2pt plus 2pt minus 2pt}
\titlespacing*{\subsection}{0pt}{4pt plus 2pt minus 1pt}{2pt plus 1pt minus 1pt}
\titlespacing*{\subsubsection}{0pt}{4pt plus 2pt minus 1pt}{2pt plus 1pt minus 1pt}

\usepackage[labelfont=bf]{caption}
\usepackage{subcaption}
\usepackage[most]{tcolorbox}
\tcbset{textmarker/.style={%
        enhanced,
        parbox=false,boxrule=0mm,boxsep=0mm,arc=1.5mm,
        outer arc=1.5mm,left=2mm,right=2mm,top=4pt,bottom=3pt,
        toptitle=1mm,bottomtitle=1mm,oversize}}
\newtcolorbox{noteBox}{textmarker,
    colback=gray!8!white}
\begin{document}

\title[Efficient Interactive LLM Serving with Proxy Model-based Sequence Length Prediction]{Efficient Interactive LLM Serving with\\Proxy Model-based Sequence Length Prediction}

% \author{\textit{Submission Category: Technical Papers}}
\author{Haoran Qiu}
\affiliation{%
  %\institution{\small \mbox{University of Illinois, Urbana-Champaign}}
  \institution{\mbox{UIUC}}
%   \city{Urbana}
%   \state{Illinois}
}
%\email{haoranq4@illinois.edu}

\author{Weichao Mao}
\affiliation{%
  %\institution{\small \mbox{University of Illinois, Urbana-Champaign}}
  \institution{\mbox{UIUC}}
%   \city{Urbana}
%   \state{Illinois}
}
%\email{weichao2@illinois.edu}

\author{Archit Patke}
\affiliation{%
  %\institution{\small \mbox{University of Illinois, Urbana-Champaign}}
  \institution{\mbox{UIUC}}
%   \city{Urbana}
%   \state{Illinois}
}
%\email{apatke@illinois.edu}

\author{Shengkun Cui}
\affiliation{%
  %\institution{\small \mbox{University of Illinois, Urbana-Champaign}}
  \institution{\mbox{UIUC}}
%   \city{Urbana}
%   \state{Illinois}
}
%\email{scui8@illinois.edu}

\author{Saurabh Jha}
\affiliation{%
  \institution{IBM Research}
%   \city{Yorktown Heights}
%   \state{New York}
}
%\email{Saurabh.Jha@ibm.com}

\author{Chen Wang}
\affiliation{%
  \institution{IBM Research}
%   \city{Yorktown Heights}
%   \state{New York}
}
%\email{chen.wang1@ibm.com}

\author{Hubertus Franke}
\affiliation{%
  \institution{IBM Research}
%   \city{Yorktown Heights}
%   \state{New York}
}
%\email{frankeh@us.ibm.com}

\author{Zbigniew Kalbarczyk}
\affiliation{%
  %\institution{\small \mbox{University of Illinois, Urbana-Champaign}}
  \institution{\mbox{UIUC}}
%   \city{Urbana}
%   \state{Illinois}
}
% \email{kalbarcz@illinois.edu}

\author{Tamer Ba\c{s}ar}
\affiliation{%
  %\institution{\small \mbox{University of Illinois, Urbana-Champaign}}
  \institution{\mbox{UIUC}}
%   \city{Urbana}
%   \state{Illinois}
}
% \email{basar1@illinois.edu}

\author{Ravishankar Iyer}
\affiliation{%
  %\institution{\small \mbox{University of Illinois, Urbana-Champaign}}
  \institution{\mbox{UIUC}}
  %\city{Urbana}
  %\state{Illinois}
  %\country{USA}
}
% \email{rkiyer@illinois.edu}
% \author[1]{Haoran Qiu}
% \author[1]{Weichao Mao}
% \author[1]{Archit Patke}
% \author[1]{Shengkun Cui}
% \author[2]{Saurabh Jha}
% \author[2]{Chen Wang}
% \author[2]{Hubertus Franke}
% \author[1]{Zbigniew Kalbarczyk}
% \author[1]{Tamer Ba\c{s}ar}
% \author[1]{Ravishankar Iyer}

% \affil[1]{University of Illinois, Urbana Champaign}
% \affil[2]{IBM Research}

\renewcommand{\shortauthors}{Haoran Qiu, Weichao Mao, et al.}

\begin{abstract}
Large language models (LLMs) have been driving a new wave of interactive AI applications across numerous domains.
However, efficiently serving LLM inference requests is challenging due to their unpredictable execution times originating from the autoregressive nature of generative models.
Existing LLM serving systems exploit first-come-first-serve (FCFS) scheduling, suffering from head-of-line blocking issues.
% SSJF addresses the non-deterministic nature of generative LLMs to enable efficient interactive inference serving.
To address the non-deterministic nature of LLMs and enable efficient interactive LLM serving, we present a speculative shortest-job-first (SSJF) scheduler that uses a light proxy model to predict LLM output sequence lengths.
Our open-source SSJF implementation does not require changes to memory management or batching strategies.
Evaluations on real-world datasets and production workload traces show that SSJF reduces average job completion times by 30.5--39.6\% and increases throughput by 2.2--3.6\texttimes{} compared to FCFS schedulers, across no batching, dynamic batching, and continuous batching settings.
\end{abstract}


% \ccsdesc[500]{Software and its engineering~Cloud computing}
% \ccsdesc[300]{Theory of computation~Scheduling algorithms}
% \ccsdesc[300]{Information systems~Language models}


\maketitle

\section{Introduction}

% \textbf{Motivation.}
Recent advancements in large language models (LLMs), trained on vast amounts of data, can engage in human-like dialogue and tackle a diverse array of tasks such as language translation, coding, and creative writing.
As LLMs continue to evolve, they drive a new wave of interactive AI applications across numerous domains.
Therefore, efficient inference serving is crucial for interactive LLM-based AI applications to provide an engaging user experience, as their interactive nature mandates low job completion times (JCT) and higher throughput for inference requests.

% \noindent
\textbf{Challenges.}
Unlike serving traditional deep neural network (DNN) model inference requests that are deterministic~\cite{gujarati2020serving} (e.g., in ResNet or BERT), serving LLM requests (e.g., GPT) faces the challenge of unpredictable execution times.
The unpredictability originates from the \textit{autoregressive} nature of such generative LLMs.
Given an input request, the \textit{output token sequence} from an LLM is generated \textit{iteratively} (i.e., token by token in each iteration)~\cite{yu2022orca}.
After an output token is generated, it is appended to the input token sequence to generate the next output token (in the next iteration).
Therefore, the execution time or the total number of iterations of serving an input request is unknown ahead of time.

Traditional DNN serving systems such as Clipper~\cite{crankshaw2017clipper}, TensorFlow-Serving~\cite{tensorflow-serving}, and Clockwork~\cite{gujarati2020serving} are designed for serving deterministic inference requests.
Since requests have no difference in execution time (for the same model on the same hardware), a first-come-first-serve (FCFS) scheduling policy is used.
State-of-the-art LLM serving systems, including Orca~\cite{yu2022orca}, vLLM~\cite{kwon2023efficient}, and AlpaServe~\cite{li2023alpaserve}, all continue to exploit an FCFS scheduling policy.
However, it is known that an FCFS request scheduling policy suffers from \textit{head-of-line blocking} issues~\cite{head-of-line-blocking,wu2023fast}, which is exacerbated for LLM inference due to the models' considerable size leading to long absolute execution times.
FastServe~\cite{wu2023fast} improves the average job completion time with a multi-level feedback queue-based approach. However, it does not address the non-determinism (which this paper addresses) and introduces additional memory management overhead to maintain intermediate states for unfinished jobs.

% \noindent
\textbf{Our Work.}
Our key observation is that a small proxy model (i.e., a fine-tuned BERT-base model) can predict the LLM verbosity well given the input query.
Based on this insight, we present the design and implementation of a \textit{speculative shortest-job-first} (SSJF) scheduler for LLM serving, using a proxy-model-based sequence length predictor for execution time estimation.
SSJF can be directly applied (1) in existing LLM serving systems with no need to change the memory or key-value cache management, and (2) in various batching settings, i.e., no batching (e.g., in \cite{li2023alpaserve}), dynamic batching (e.g., in \cite{triton}), and continuous (iteration) batching (e.g., in \cite{yu2022orca}).

% \noindent
\textbf{Results.}
We evaluate \xxx with popular open-source LLMs and a real-world LLM conversation dataset~\cite{zheng2023lmsys} on NVIDIA V100 GPUs, driven by production workload traces.
We evaluate in various batching settings: no batching, dynamic batching, and continuous batching.
Evaluation results show that \xxx reduces the average JCT by 30.5--39.6\% and increases the throughput by 2.2--3.6\texttimes{} at varying request rates and burstinesses compared to FCFS schedulers.

\textbf{Contributions.}
Our main contributions are:
\begin{itemize}
    \item A speculative request scheduler with a light proxy model (i.e., \xxx) that addresses non-determinism in generative LLMs.
    \item An open-source implementation of \xxx and evaluation on real-world model datasets and production traces.
    \item We discuss other potential use cases of proxy models in LLM serving in addition to request scheduling (\cref{sec:discussion}).
\end{itemize}
\section{Background and Motivation}
\label{sec:background}

\noindent
\textbf{Serving Interactive LLM Applications.}
Large language models (LLMs) (e.g., GPT) are Transformer-based generative models.
LLMs have enabled many interactive AI applications such as Chatbots~\cite{chatgpt}, AI pair programmers~\cite{copilot}, and search~\cite{bard}.
LLMs are trained to generate each output token sequence in an \textit{autoregressive} manner.
Therefore, to process a request to generative models, multiple \textit{iterations} of the model have to be run; each iteration generates a single output token, which is then appended to the original input in the following iteration (except for the termination token \texttt{<EOS>}).

The autoregressive patterns lead to non-deterministic execution times when serving LLM inference workloads.
We analyze the output token length distribution in the LMSYS-Chat-1M dataset~\cite{zheng2023lmsys} which contains one million real-world conversations with 25 state-of-the-art LLMs.
The output token length varies significantly for each model instance.
The p95/p50 of the output token length for each model varies from 1.7 (\texttt{claude-1}) to 20.5 (\texttt{llama-13b}).
The output token length ($N$) dominates the execution time ($T$) of a request because $T = C + K * N$, where $K$ is the latency to generate one token and $C$ is the model-serving system's overhead including DNS lookups, proxies, queueing, and input tokenization.
$K$ depends on model optimization techniques (e.g., quantization) and execution environment (e.g., hardware), which are the same for all inputs.

\noindent
\textbf{Scheduling and Batching.}
% Talk about existing techniques for scheduling and batching in model serving in general
ML model serving systems serve as an abstraction wrapping the underlying execution engine to queue the arriving requests, dispatch jobs to available computing devices such as GPUs (i.e., \textit{scheduling}), and return the results to the end users.
Model serving systems typically \textit{batch} requests to increase hardware utilization (by leveraging parallel computing units in hardware accelerators) and system throughput.
The batched input requests are concatenated and fed into the model as a whole.
However, the non-deterministic execution times of serving autoregressive models pose challenges in both scheduling and batching.

First, a first-come-first-serve (FCFS) scheduling policy for serving LLMs may suffer from \textit{head-of-line blocking} issues, leading to potentially high waiting times and thus long job completion times (JCT).
Traditional DNN serving systems treat requests coming to the same model instance equally, leveraging FCFS scheduling policies, because requests for the same model instance have deterministic execution times (i.e., when serving ResNet and BERT models)~\cite{crankshaw2017clipper,tensorflow-serving,inferline,gujarati2020serving}. 
FastServe~\cite{wu2023fast} uses a multi-level feedback queue-based scheduling approach to improve the average JCT but it does not address the non-determinism nature.

Second, batching requests of various output token sequence lengths can lead to low hardware utilization as early finished requests are waiting for the unfinished longer requests in the same batch.
Orca~\cite{yu2022orca} proposes iterative scheduling and continuous batching to re-dispatch a new batch of requests when there is an early finished request in the last batch.
However, Orca leverages FCFS so that there could still be head-of-line blocking and long JCTs.
% In addition, there can be frequent context switch overhead as early finished request is frequently dropped out of the batch.
\section{Method}
\label{sec:method}

% \subsection{Overview}
% \label{sec:method:overview}

\begin{figure}[!t]
    \centering
    \includegraphics[width=0.985\linewidth]{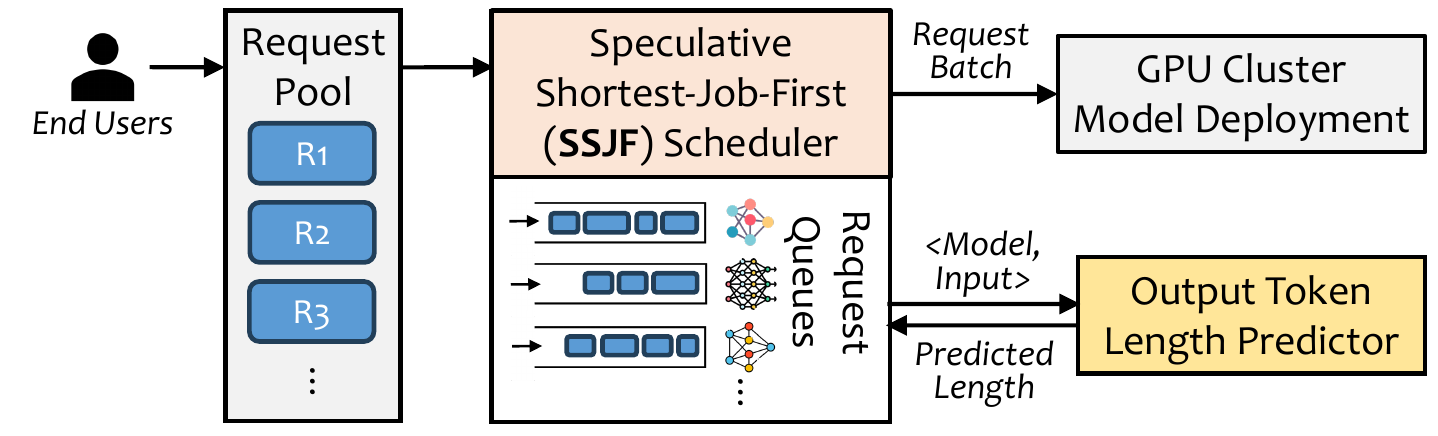}
    % \vspace{-5pt}
    \caption{\xxx overview.}
    \label{fig:scheduler-arch}
\end{figure}

Input requests from users are dispatched to corresponding models at runtime.
% For models with multiple replicas, the request is dispatched to the one with the shortest queue length (i.e., the sum of the job length estimates).
For each model instance, a request queue is maintained, and \xxx runs a \textit{speculative} shortest-job-first scheduler to decide the request execution order.
SJF alleviates the non-determinism of generative models and the head-of-line blocking issue in FCFS scheduling.
Since we do not have the ground truth execution time of each request (required by SJF), as shown in \cref{fig:scheduler-arch}, \xxx relies on the prediction from an \textit{output token length predictor}.
The prediction is used to estimate the job execution time in SJF as the output token length dominates the execution time (linear relation) as described in \cref{sec:background}.
The predictor relies on a lightweight proxy model, i.e., a BERT-base model in our case.

\subsection{Proxy Model for Output Length Prediction}
\label{sec:method:proxy}

\begin{figure}[!t]
    \centering
    \includegraphics[width=0.985\linewidth]{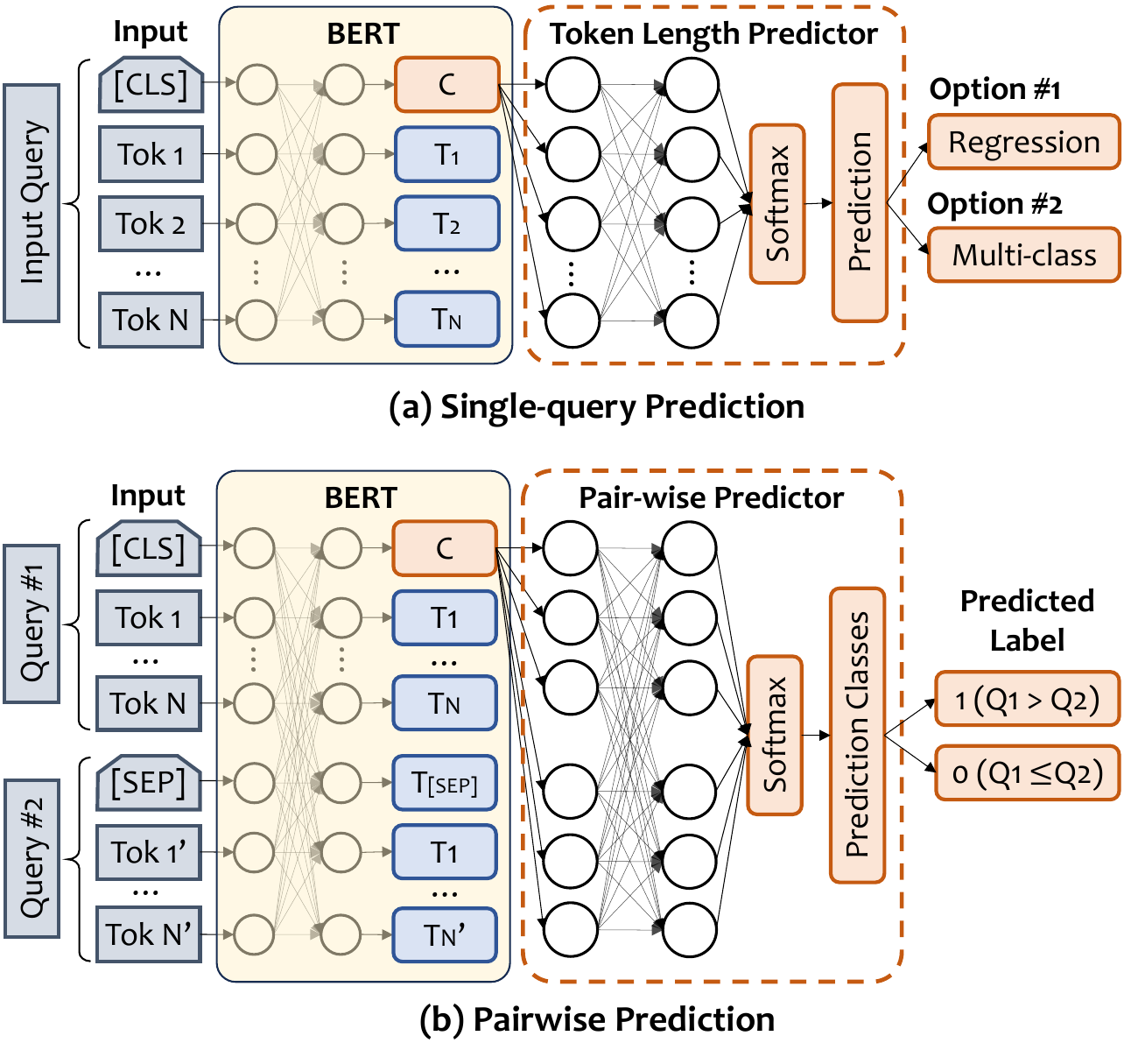}
    \caption{Output length predictor architecture.}
    \vspace{5pt}
    \label{fig:predictor-arch}
\end{figure}

\noindent
\textbf{Problem Formulation.}
We formulate the output token length prediction for each query as a regression problem (option \#1) or a multi-class classification problem (option \#2, similar to the formulation in S$^3$~\cite{jin2023s}), as shown in \cref{fig:predictor-arch}(a).
In the regression task, the prediction is the absolute output token length while in classification, the prediction is a category of percentile values.
For example, in binary-class classification, two categories could be [0, p50) and [p50, max], where p50 is the median output token length based on historical LLM serving records.

An alternative formulation is that we can have a pairwise predictor (as shown in \cref{fig:predictor-arch}(b)) whose input are two LLM queries (Q1 and Q2) and the prediction is 1 if Q1's output is longer than Q2's and 0 otherwise.
Given such a pairwise predictor, the SJF scheduler can insert any new arrival request to the sorted queue (based on output length) based on the pairwise prediction of the new request and any existing requests in the queue.
However, this alternative approach can incur $O(log N)$ proxy model inference times (where $N$ is the maximum queue length) compared to $O(1)$ in SSJF's single-query prediction approach.
In addition, evaluation results show that SJF with pairwise prediction has only $\sim 3$\% improvement in JCT compared to FCFS (SSJF has $>10\times$ higher improvement).
Therefore, we do not proceed with this approach.

\begin{figure}[!htb]
    \centering
    \includegraphics[width=0.8\linewidth]{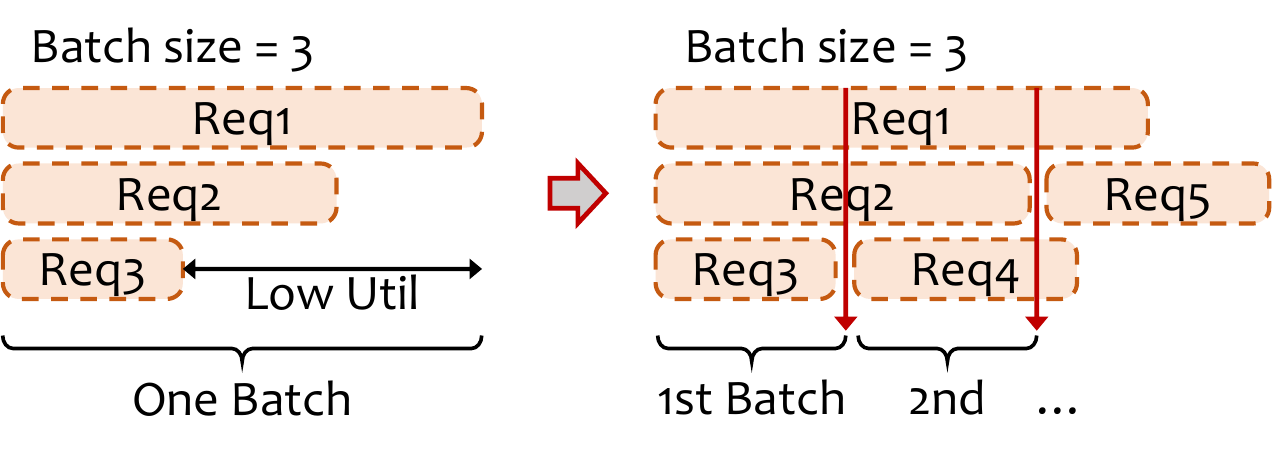}
    % \vspace{-5pt}
    \caption{Batching vs. Continuous batching.}
    \vspace{3pt}
    \label{fig:batching}
\end{figure}

\noindent
\textbf{Predictor Architecture.}
% Talk about the proxy model and the predictor architecture
As shown in \cref{fig:predictor-arch}(a), we use BERT-base as a proxy model and append a two-layer fully connected neural network with a softmax layer to BERT-base as the output token length predictor.
Specifically, we take the last layer hidden state of the first token (i.e., \texttt{CLS}) from the BERT output.
As suggested by the BERT paper~\cite{bert}, the final hidden state corresponding to \texttt{CLS} is used as the aggregate sequence representation for classification tasks.

\noindent
\textbf{Fine-tuning Proxy Model.}
% Talk about the fine-tuning process
Starting from the pre-trained BERT-base model provided by~\cite{bert}, we fine-tune the weights (together with the additional two-layer prediction network) on the LMSYS-Chat-1M training data~\cite{zheng2023lmsys}. 
Pre-training on a large corpus of unlabeled text gives the proxy model some basic understanding of human languages, while the fine-tuning step allows the proxy model to learn the \textit{task-specific} knowledge (in our case, the output token length of a generative model given the input query).
Our fine-tuning has two phases: In the first phase, the parameters of both BERT-base and the predictor are tuned, while in the second phase, we fix the weights of BERT-base and only update the predictor's parameters. 
Such a two-phase regime turns out to achieve a good balance between prediction accuracy and training efficiency.

In the regression task formulation (option \#1), we can use either L1 loss or MSE loss.
In the multi-class classification formulation (option \#2), we can either use the standard cross-entropy loss or adopt an ordinal classification training approach \cite{winship1984regression}. 
% Since these classes have an inherent order, we face an intermediate problem between regression and classification. 
For example, in five-class classification, we convert the classes into 5 integers $\{0,1,2,3,4\}$ and let our predictor output a real value as the prediction.
We can then apply L1 or MSE loss to minimize the distance between the real-valued predictions and the integer-valued class labels.
During inference, we round the prediction value to the nearest integer in $\{0,\dots,4\}$.
As shown in \cref{sec:eval}, the regression formulation with L1 loss (i.e., REG-L1) achieves the best prediction accuracy and scheduling performance.

\noindent
\textbf{Support for Multi-round LLM Conversations.}
In addition to general LLM inferences, \xxx also supports interactive LLM inference that happens in rounds.
Previous rounds of user input queries and LLM responses are usually provided as \textit{context} to the current-round input query.
However, the input limit to BERT-base as a proxy model is 512 tokens, which can be much less than the context window.
% Talk about how to support multi-round conversations given the input limit for BERT is 512
To support multi-round conversations, we use a simple strategy of concatenation and truncation.
Specifically, we concatenate all the previous-round user prompts (P1, P2, P3, etc.), LLM responses (R1, R2, R3, etc.), and the current-round user prompt to form a context-augmented input query for the current round of conversation.
Then, we retain only the last 512 tokens of the concatenated query, dropping the earlier portions.
By concatenating the previous conversation history and truncating it if necessary, this approach allows a proxy model to effectively handle multi-round conversations while staying within the input length limitation.

\subsection{Scheduling and Batching}
\label{sec:method:scheduling}

% Talk about how the output length prediction is used for scheduling and batching

With predictions of the output token length of each incoming inference request, SSJF schedules requests with shorter predicted lengths to run first.
SSJF supports three modes of batching: no batching, dynamic batching, and continuous batching.
In the no-batching mode, the request is served one by one according to the execution order determined by SSJF.
In the dynamic-batching mode, the scheduler waits for \texttt{batch$\_$wait$\_$timeout} or until \texttt{max$\_$batch$\_$size} requests in each batch have been filled up whose order is determined by SSJF.
In the continuous-batching mode (first proposed by Orca~\cite{yu2022orca}), requests are batched at the iteration level and the early-finished request is substituted by the next request in the waiting queue (ordered by SSJF).
The difference between (dynamic) batching and continuous batching is illustrated in \cref{fig:batching}.
\begin{table}[!b]
    \centering
    \caption{Models used in experiments. \textit{The latency is measured for a query with a sequence length of 512 on a single GPU.}}
    % \vspace{-5pt}
    \resizebox{0.8\linewidth}{!}{%
    \begin{tabular}{llll}
    \toprule
    Model & \# of Params & Size & Latency\\
    \midrule
    % OPT-125m & 125 M & 0.5 GB & 134 ms\\
    OPT-1.3b & 1.3 B & 5.0 GB & 1243 ms\\
    OPT-2.7b & 2.7 B & 10.4 GB & 2351 ms\\
    % GPT2 & 137 M & 0.6 GB & 144 ms\\
    % GPT2-medium & 355 M & 1.5 GB & 404 ms\\
    GPT2-large & 774 M & 3.3 GB & 832 ms\\
    GPT2-xl & 1.5 B & 6.4 GB & 1602 ms\\
    CodeGen-350m & 350 M & 1.3 GB & 357 ms\\
    CodeGen-2b & 2.0 B & 8.0 GB & 2507 ms\\
    % Bloom-560m & 560 M & 2.0 GB & 298 ms\\
    Bloom-1b1 & 1.1 B & 4.0 GB & 523 ms\\
    % Bloom-1b7 & 1.7 B & 6.0 GB & 790 ms\\
    Bloom-3b & 3.0 B & 11.0 GB & 1293 ms\\
    % Switch-base-8 & 460 M & 1.2 GB & 316 ms\\
    Switch-base-16 & 920 M & 2.4 GB & 348 ms\\
    Switch-base-32 & 1.8 B & 4.8 GB & 402 ms\\
    \bottomrule
    \end{tabular}%
    }
    \label{table:model-details}
\end{table}

\section{Evaluation}
\label{sec:eval}

\subsection{Experiment Setup}
\label{sec:eval:setup}

\begin{figure*}[!t]
    \captionsetup[subfigure]{aboveskip=-2pt,belowskip=0pt}
    \begin{subfigure}[b]{0.34\linewidth}
        \centering
        {\includegraphics[width=\linewidth]{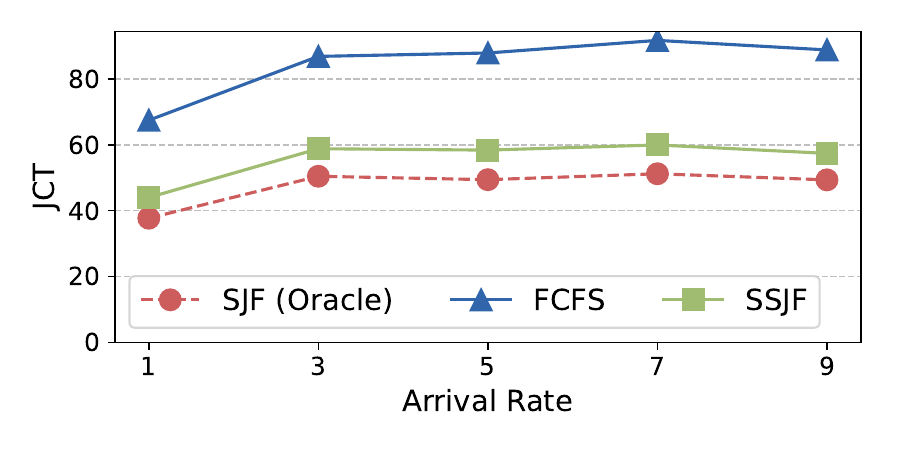}}
        \caption{No batching.}
    \end{subfigure}%
    \hfill%
    \hspace{-20pt}
    \begin{subfigure}[b]{0.34\linewidth}
        \centering
        {\includegraphics[width=\linewidth]{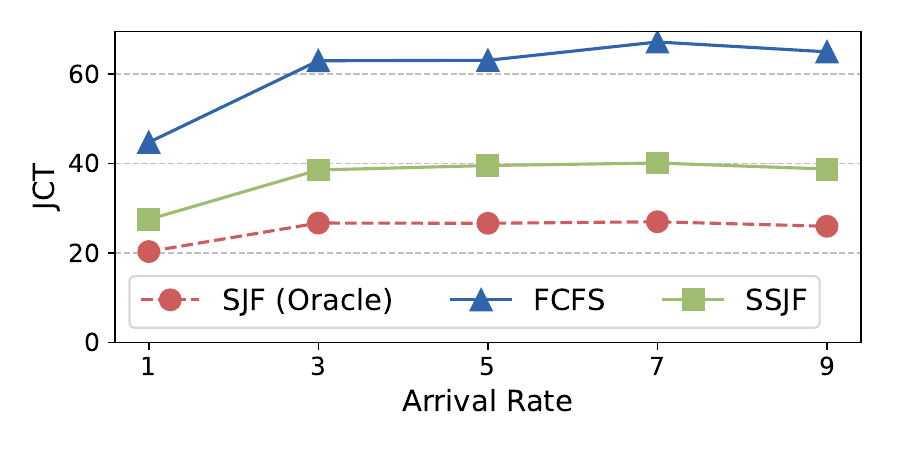}}
        \caption{Dynamic batching.}
    \end{subfigure}%
    \hfill%
    \hspace{-20pt}
    \begin{subfigure}[b]{0.34\linewidth}
        \centering
        {\includegraphics[width=\linewidth]{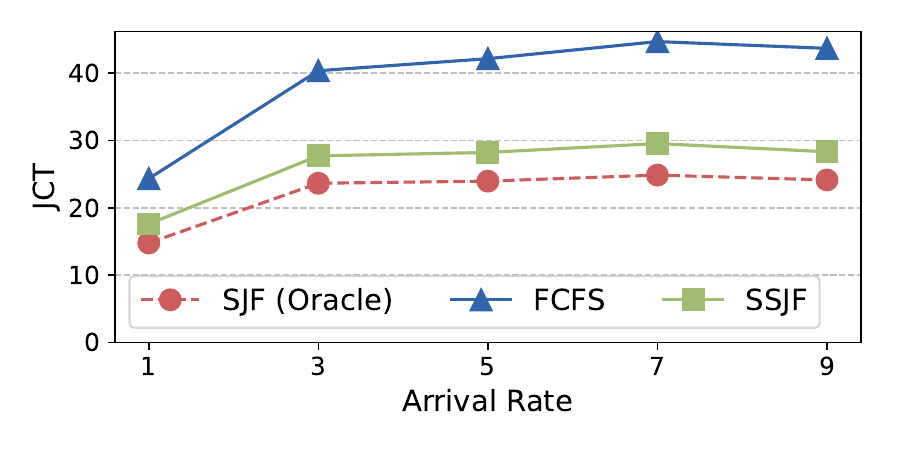}}
        \caption{Continuous batching.}
    \end{subfigure}%
    \caption{Job completion time (JCT) with varying rates.}
    \label{fig:jct-rate}
\end{figure*}

\begin{figure*}[!t]
    \captionsetup[subfigure]{aboveskip=-2pt,belowskip=0pt}
    \begin{subfigure}[b]{0.34\linewidth}
        \centering
        {\includegraphics[width=\linewidth]{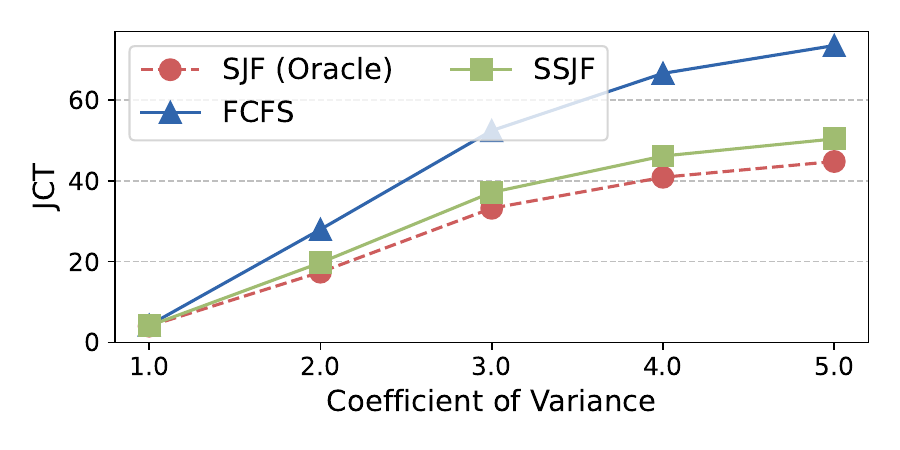}}
        \caption{No batching.}
    \end{subfigure}%
    \hfill%
    \hspace{-20pt}
    \begin{subfigure}[b]{0.34\linewidth}
        \centering
        {\includegraphics[width=\linewidth]{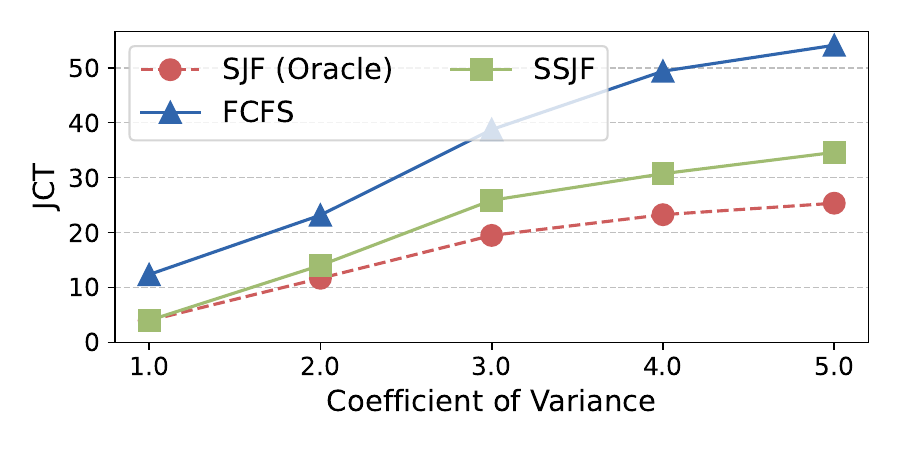}}
        \caption{Dynamic batching.}
    \end{subfigure}%
    \hfill%
    \hspace{-20pt}
    \begin{subfigure}[b]{0.34\linewidth}
        \centering
        {\includegraphics[width=\linewidth]{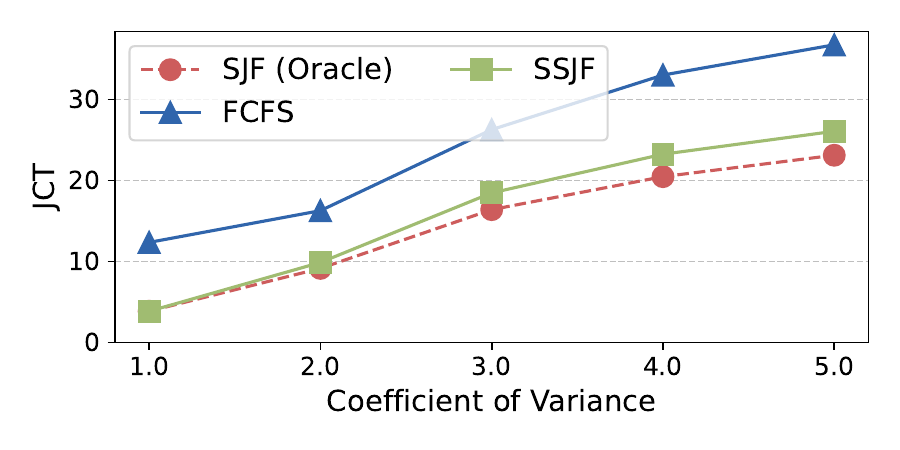}}
        \caption{Continuous batching.}
    \end{subfigure}%
    \caption{Job completion time (JCT) with varying burstiness.}
    \label{fig:jct-cv}
\end{figure*}

\begin{figure*}[!t]
    \captionsetup[subfigure]{aboveskip=-2pt,belowskip=0pt}
    \begin{subfigure}[b]{0.34\linewidth}
        \centering
        {\includegraphics[width=\linewidth]{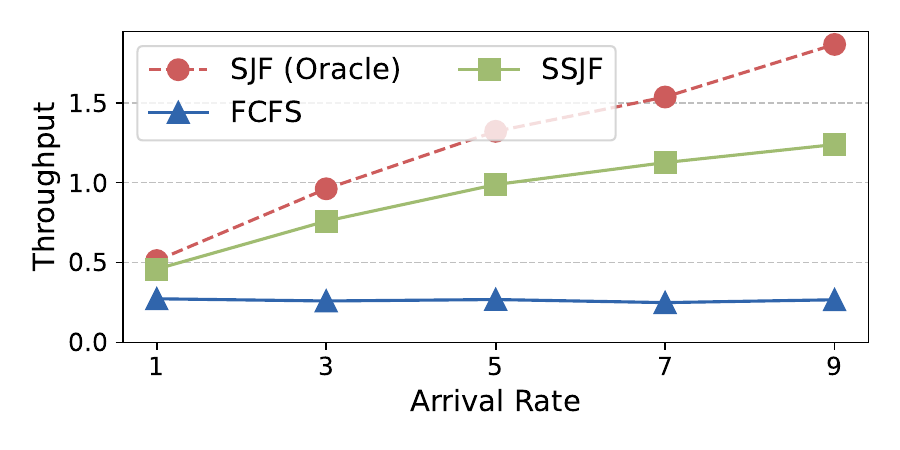}}
        \caption{No batching.}
    \end{subfigure}%
    \hfill%
    \hspace{-20pt}
    \begin{subfigure}[b]{0.34\linewidth}
        \centering
        {\includegraphics[width=\linewidth]{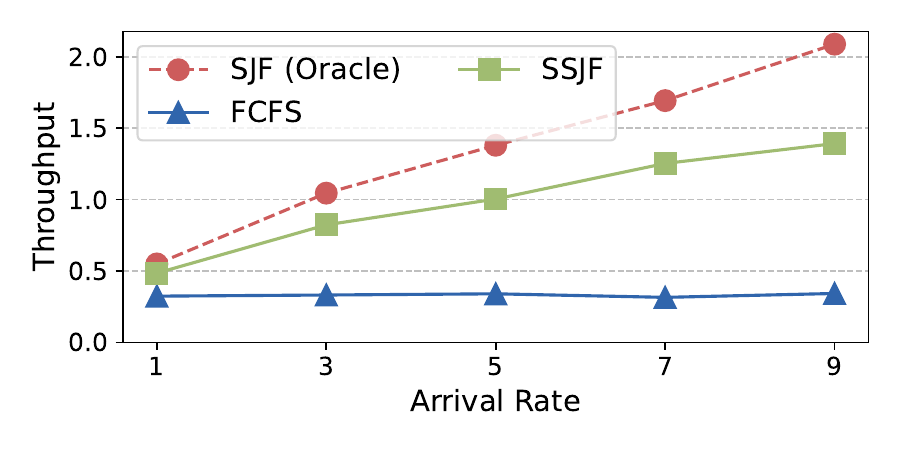}}
        \caption{Dynamic batching.}
    \end{subfigure}%
    \hfill%
    \hspace{-20pt}
    \begin{subfigure}[b]{0.34\linewidth}
        \centering
        {\includegraphics[width=\linewidth]{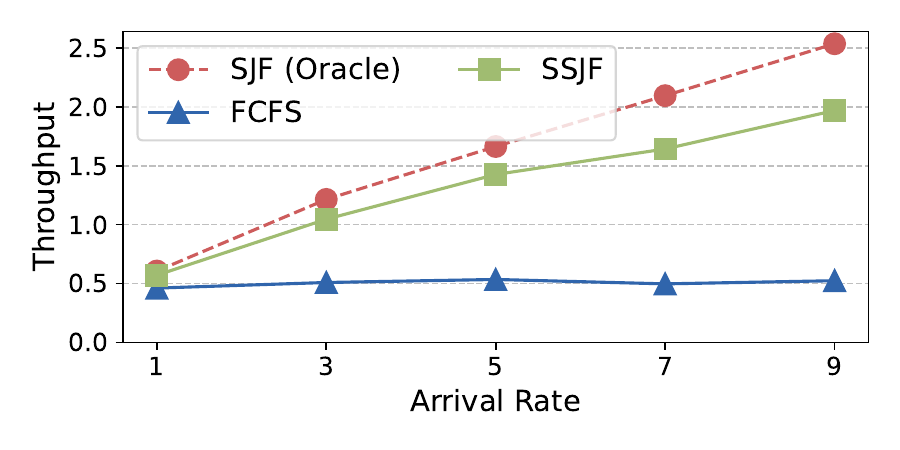}}
        \caption{Continuous batching.}
    \end{subfigure}%
    \caption{Throughput with varying rates.}
    \label{fig:throughput-rate}
\end{figure*}

\begin{figure*}[!t]
    \captionsetup[subfigure]{aboveskip=-2pt,belowskip=0pt}
    \begin{subfigure}[b]{0.34\linewidth}
        \centering
        {\includegraphics[width=\linewidth]{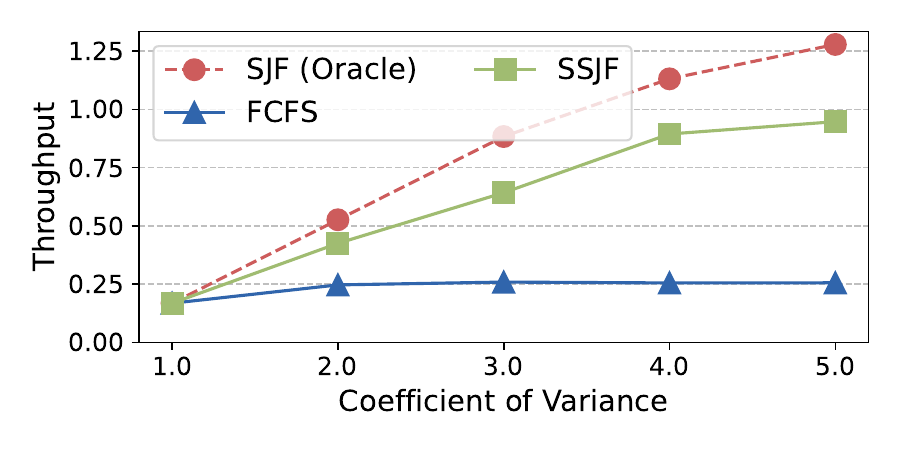}}
        \caption{No batching.}
    \end{subfigure}%
    \hfill%
    \hspace{-20pt}
    \begin{subfigure}[b]{0.34\linewidth}
        \centering
        {\includegraphics[width=\linewidth]{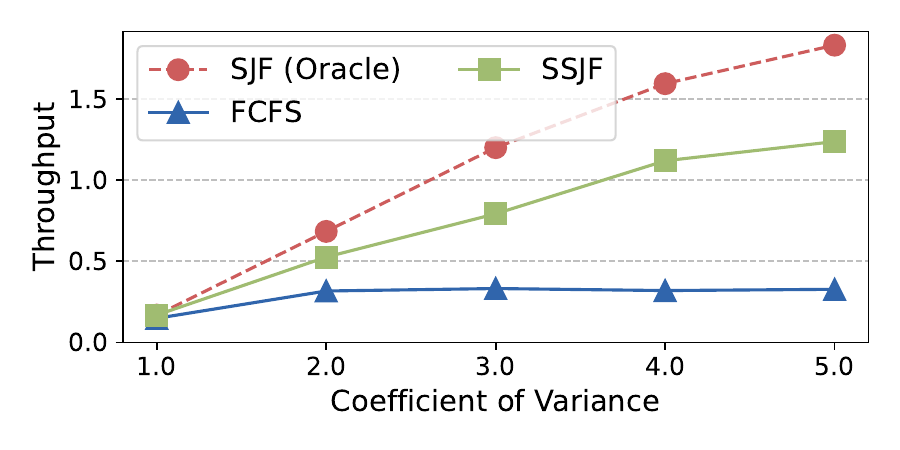}}
        \caption{Dynamic batching.}
    \end{subfigure}%
    \hfill%
    \hspace{-20pt}
    \begin{subfigure}[b]{0.34\linewidth}
        \centering
        {\includegraphics[width=\linewidth]{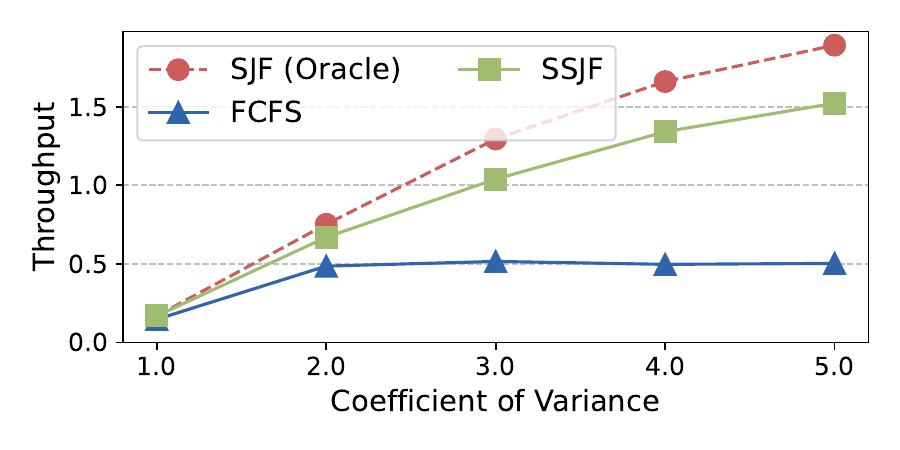}}
        \caption{Continuous batching.}
    \end{subfigure}%
    \caption{Throughput with varying burstiness.}
    \label{fig:throughput-cv}
\end{figure*}

\textbf{Models.}
Since \xxx is a general ML model-serving framework, we consider traditional non-Transformer models, Transformers, and Transformer-based generative models for evaluation.
% In practice, large models are usually pre-trained in multiple variants with different amounts of model weights.
For each model family, we select several most commonly used model sizes and variants (to mimic different fine-tuned versions) for experimentation.
\cref{table:model-details} provides details about model sizes and inference latency on testbed GPUs.

\noindent
\textbf{Workloads.}
% 17298/1753078 = 1%
% There does not exist an open-source production ML inference trace to the best of our knowledge.
We use the Microsoft Azure function traces~\cite{zhang2021faster} to drive the inference workloads.
For language model inputs, we use the LMSYS-Chat-1M dataset~\cite{zheng2023lmsys} which contains one million real-world conversations with 25 state-of-the-art LLMs.

\noindent
\textbf{Testbed.}
We deploy \xxx on an IBM Cloud \texttt{gx2-16x128x2v100} instance with 2 NVIDIA Tesla V100 (16GB) GPUs.
Each GPU supports a maximum Streaming Multiprocessor (SM) frequency of 1380 MHz and a minimum of 200 MHz.

\subsection{Scheduling Performance}
\label{sec:eval:scheduling}

We evaluate the scheduling performance regarding the job completion time (JCT) and the serving throughput to understand if prediction accuracy is sufficient.
As mentioned in \cref{sec:method:proxy}, we choose to use regression with L1 loss in the predictor because of its higher accuracy in our ablation study (as shown in the next section \cref{sec:eval:ablation}).
Our comparison baselines include (1) first-come-first-serve (FCFS), which is the default scheduler in state-of-the-art model-serving frameworks such as Orca~\cite{yu2022orca} and vLLM~\cite{kwon2023efficient}, and (2) shortest-job-first (SJF) with the actual output token length (i.e., oracle).

\noindent
\textbf{Request Serving JCT.}
% Talk about the end-to-end evaluation regarding the scheduling latency, compared with FCFS, without cache, and Oracle
% The experiments are under three settings: no batching, dynamic batching, and continuous batching
We evaluate \xxx and the comparison baselines in three batching settings: (1) no batching, (2) dynamic batching, and (3) continuous batching (max batch size is set to 4 for both batching cases).
\cref{fig:jct-rate} and \cref{fig:jct-cv} show the average LLM serving JCT with varying request arrival rates and burstiness.
With varying request rates, \xxx reduces JCT by 34.5\%, 39.6\%, and 33.2\% compared to the FCFS scheduler, under three batching settings respectively.
In comparison, the oracle SJF scheduler reduces the JCT by 43.7\%, 58.2\%, and 43.0\%.
Under different burstinesses, \xxx reduces JCT by 30.5\%, 39.0\%, 35.0\% compared to FCFS, while the oracle SJF reduces the JCT by 37.6\%, 52.9\%, and 41.5\%.

\noindent
\textbf{Request Serving Throughput.}
% Talk about the end-to-end evaluation regarding the scheduling throughput, compared with FCFS, without cache, and Oracle
Under the same experimental settings, we present the throughput results in \cref{fig:throughput-rate} and \cref{fig:throughput-cv}.
With varying request rates, \xxx increases the throughput by 3.6\texttimes{}, 3.0\texttimes{}, and 2.8\texttimes{} in the no batching, dynamic batching, and continuous batching cases, respectively.
In comparison, the oracle SJF increases the throughput by 4.7\texttimes{}, 4.1\texttimes{}, and 3.2\texttimes{}.
Under different burstinesses, \xxx increases the throughput by 2.6\texttimes{}, 2.6\texttimes{}, and 2.2\texttimes{} while the oracle SJF increases the throughput by 3.4\texttimes{}, 3.8\texttimes{}, and 2.7\texttimes{}.

\begin{figure}[!t]
    \centering
    \includegraphics[width=\linewidth]{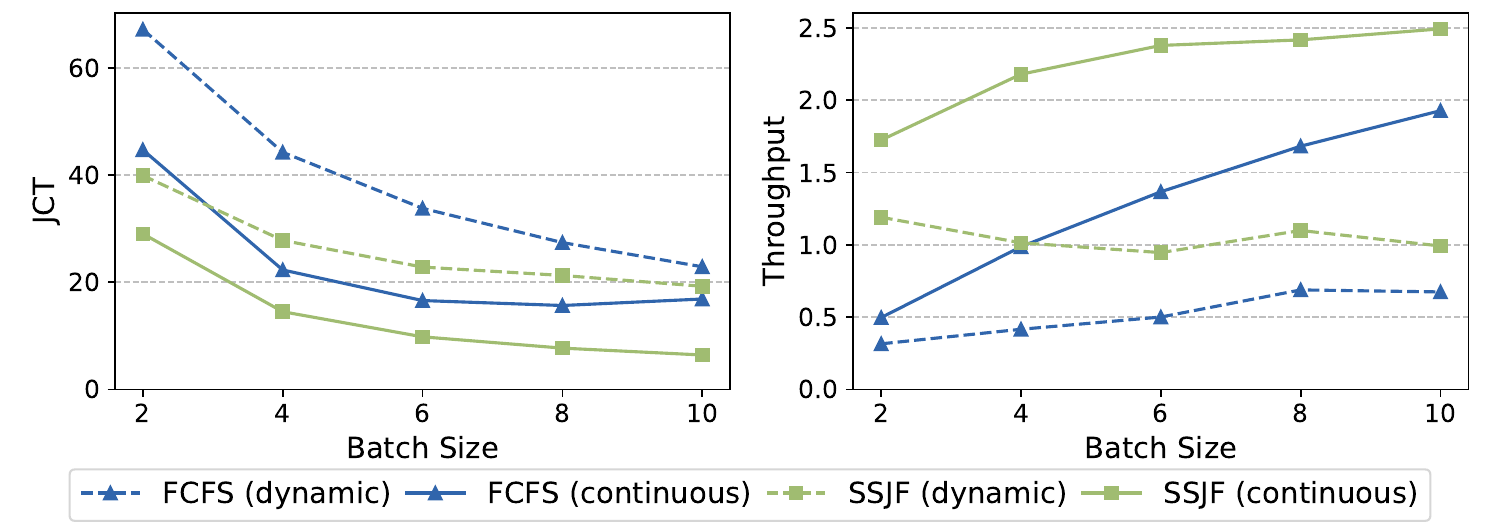}
    % \vspace{2pt}
    \caption{Benefits of SSJF at varying batch sizes.}
    \label{fig:batch-sizes}
\end{figure}

\noindent
\textbf{At Varying Batch Sizes.}
We repeat the experiments while increasing the batch sizes. Results are shown in \cref{fig:batch-sizes}.
Continuous batching always has less JCT and higher throughput than dynamic batching (same results as shown in \cite{yu2022orca}).
At varying batch sizes, \xxx's scheduler performance is better than FCFS's but the improvements in JCT and throughput that \xxx brings decrease as the batch size increases.
This is because the larger the batch, the less difference there is when reordering the queue.

\noindent
\textbf{At Different Conversation Rounds.}
We evaluate the effectiveness of \xxx's support for multi-round LLM conversations.
Results are shown in \cref{fig:rounds}.
At different conversation rounds, the improvement of JCT (compared to FCFS) varies from 33.1\% (at round \#5) to 38.9\% (at round \#1); The improvement of throughput varies from 1.67\texttimes{} (at round \#5) to 2.22\texttimes{} (at round \#1).
% Round #0 JCT: 38.92% Throughput: 2.22x
% Round #1 JCT: 34.16% Throughput: 1.78x
% Round #2 JCT: 33.84% Throughput: 1.73x
% Round #3 JCT: 34.54% Throughput: 1.74x
% Round #4 JCT: 33.07% Throughput: 1.67x
% Round #all JCT: 39.14% Throughput: 2.21x

\noindent
\textbf{Predictor Overhead.}
We compare the predictor latency overhead (i.e., proxy model inference) and LLM request execution time.
The average model execution time is 9.8 s, greater than the average predictor latency of 7.6 ms (0.02\% of the total latency).
The maximum predictor latency is 20.2 ms, which is less than the minimum model execution time of 120 ms.
% The p5th model execution time is 360 ms.
Therefore, we conclude that the predictor introduces negligible overhead to the end-to-end LLM-serving latency, which supports the design and benefits of using a lightweight proxy model in \xxx.

\subsection{LLM Output Length Prediction}
\label{sec:eval:ablation}

As mentioned in \cref{sec:method:proxy}, we ended up using regression with L1 loss in the predictor because of its higher accuracy in our ablation study.
In this section, we evaluate the predictor's accuracy and present several ablation studies in the prediction method and training method.

\begin{figure}[!t]
    \centering
    % \vspace{-5pt}
    \includegraphics[width=0.85\linewidth]{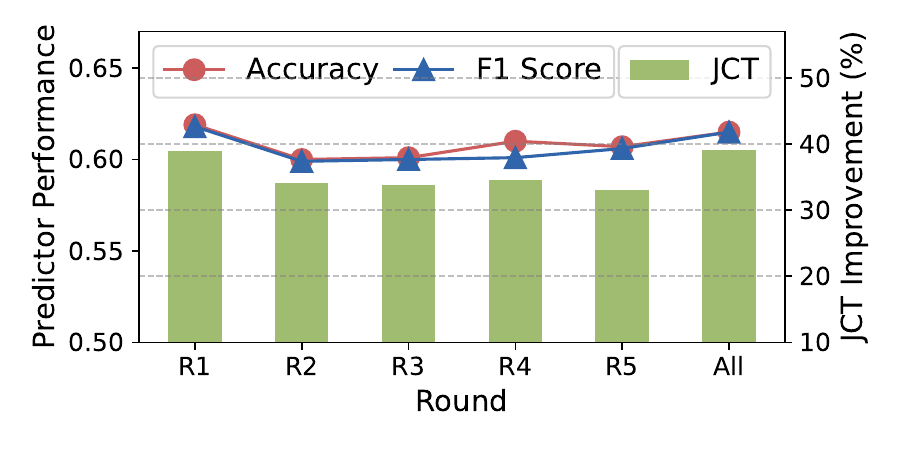}
    \vspace{-5pt}
    \caption{SSJF evaluation across conversation rounds.}
    \label{fig:rounds}
\end{figure}

\noindent
\textbf{Prediction Accuracy on All Models.}
% Talk about the prediction accuracy of all models
We evaluate the output length predictor on the LMSYS-Chat-1M dataset~\cite{zheng2023lmsys}. 
The predictor achieves an average accuracy of 0.615 across all models (translating to a five-class classification whose random guess accuracy is 0.2).
As shown in \cref{fig:rounds}, at different rounds, the prediction accuracy ranges from 0.60 (round \#2) to 0.619 (round \#1) and the prediction F1 scores range from 0.599 (round \#2) to 0.618 (round \#1).

\noindent
\textbf{Ablation Study.}
% Talk about the prediction accuracy + scheduling performance of different predictors: (1) binary, (2) 5-class, and (3) regression.
% Talk about the prediction accuracy + scheduling performance of different predictors: (1) standard classification, (2) ordinal classification with L1 loss, and (3) ordinal classification with MSE loss.
In Table~\ref{table:predictor_eval}, we compare our predictor with multiple alternative approaches, including binary classification, ordinal multi-class classification with MSE/L1 loss, (standard) classification with cross-entropy loss, and regression with MSE loss.
The regression approach with L1 loss used by \xxx's predictor achieves the highest prediction accuracy, F1 score, average JCT improvement, and throughput improvement.

\begin{table}[!tb]
\def\arraystretch{1.05}
\centering
\caption{Output token length predictor evaluation results. \textit{\xxx adopts regression with L1 loss, i.e., REG (L1), given its better accuracy and scheduling performance improvement.}}
% \vspace{-5pt}
\label{table:predictor_eval}
\resizebox{0.98\linewidth}{!}{%
\begin{tabular}{rcccccc}
\toprule
 Metrics& Bin. CLS & \begin{tabular}[]{@{}c@{}}Ord. CLS\\(MSE) \end{tabular} & \begin{tabular}[]{@{}c@{}}Ord. CLS\\(L1)\end{tabular} & CLS & \begin{tabular}[]{@{}c@{}}REG\\(MSE)\end{tabular} & \begin{tabular}[]{@{}c@{}}REG\\(L1)\end{tabular} \\ \midrule
 % single-round
 % Accuracy $\uparrow$ & N/A & 0.515 & 0.575 & 0.590 & 0.591 & \textbf{0.615} \\
 % F1 Score $\uparrow$ & N/A & 0.485 & 0.587 & 0.602 & 0.595 & \textbf{0.615} \\ % weighted
 % F1 Score $\uparrow$ & N/A & 0.377 & 0.461 & \textbf{0.498} & 0.464 & 0.480 \\ % macro - harmonious average
 % multi-round
 Accuracy $\uparrow$ & N/A & 0.477 & 0.580 & 0.592 & 0.601 & \textbf{0.615} \\
 F1 Score $\uparrow$ & N/A & 0.450 & 0.576 & 0.590 & 0.602 & \textbf{0.615} \\ % weighted
 \midrule
 JCT Reduction $\uparrow$ & 30.8\% & 38.0\% & 35.6\% & 35.8\% & 38.6\% & \textbf{39.1\%} \\
 Throughput $\uparrow$ & 1.59\texttimes{} & 2.15\texttimes{} & 1.99\texttimes{} & 1.98\texttimes{} & 2.19\texttimes{} & \textbf{2.21\texttimes{}} \\
 % L1 Error $\downarrow$ & 0.7587 & 0.7480 & N/A & 0.6077 & \textbf{0.5557} \\
 % MSE $\downarrow$ & 0.9859 & 1.0857 & N/A & \textbf{0.8077} & 0.8996 \\
\bottomrule
\end{tabular}
}
\end{table}

\noindent
\textbf{Note on Data Processing.}
Since there are empty responses and randomly truncated responses (which only happens to a subset of responses with length >512) in the LMSYS-Chat-1M dataset~\cite{zheng2023lmsys} due to unknown reasons, we filtered the dataset with response length (1, 512).
Before filtering the dataset, multi-class classification with MSE error resulted in a prediction accuracy of 0.42, which outperformed all other alternative approaches in the predictor.
Therefore, choosing the right prediction method (i.e., regression or classification, the number of classes) can be dataset- or model-dependent.
\section{Related Work}

\noindent
\textbf{LLM Output Length Prediction.}
% In machine translation, Yan et al.~\cite{yang-etal-2020-predicting} propose a combination of a convolutional layer, a max-pooling layer, and linear layers, to obtain a single vector representation of input tokens, which is then fed into a softmax classifier over possible target lengths.
% hu2024inference, zheng2024response, jin2023s
$S^{3}$~\cite{jin2023s} proposes a multi-class classifier based on DistilBERT (a Transformer model trained by distilling BERT base) to classify the output token length of an LLM input query to ten evenly divided length buckets.
Similarly, TetriInfer~\cite{hu2024inference} leverages a classification model
(i.e., a {125M OPT model}) for length prediction.
However, $S^{3}$ does not support multi-round conversations and is only trained using a 52K question-answering dataset ($\sim$2\% of the LMSYS-Chat-1M data that we use).
Our experiments on the larger dataset show that, compared to multi-class classification, the regression method used in SSJF further improves scheduling JCT and throughput.
Lastly, Zheng et al.~\cite{zheng2024response} propose to prompt the LLM itself for length prediction. However, this approach can introduce bias (due to modification in the same user prompt) and higher overhead compared to a proxy model-based approach.

\noindent
\textbf{Prediction-based SJF.}
Outside the ML model serving domain, there are examples of prediction-based SJF schedulers proposed in big data compute clusters~\cite{276962,6868256} and HPC clusters~\cite{naghshnejad2020hybrid,prep}.
For instance, SLearn~\cite{276962} estimates the job runtime based on the history of similar jobs and uses the prediction for SJF scheduling on Hadoop.
PREP~\cite{prep} leverages support vector regression to estimate HPC job runtime based on features such as job running state, resource allocation, and user/group ID.
Compared to HPC systems or Hadoop-like big data compute clusters, the proxy-model-based sequence length prediction method proposed by this paper addresses the unique non-deterministic nature of autoregressive models to enable efficient interactive LLM inference serving.

\section{Discussion and Future Work}
\label{sec:discussion}
\noindent
\textbf{Starvation Handling and Preemption.}
To avoid starvation in SJF scheduling, one can adopt aging~\cite{aging} to promote jobs with long waiting times.
Preemption could also help correct previous suboptimal decisions with a least-slack-time-first policy.
However, \xxx does not exploit preemption due to its added complexity in context switch~\cite{bai2020pipeswitch} and memory management~\cite{kwon2023efficient}.

\noindent
\textbf{Use of Semantic Cache.}
% Talk about how we use semantic cache to increase the prediction accuracy
To improve the output token length prediction accuracy, a simple \textit{semantic cache} (e.g., GPTCache~\cite{bang2023gptcache}) can be used to cache the ground truth output token sequence length of hot inputs.
Instead of going through the predictor to get an estimated output length prediction, the scheduler retrieves the cached length for the input that triggers a cache hit (i.e., there are semantically similar input queries in the cache).
% Caching has been commonly used to reduce frequent and computationally expensive data accesses.
% A semantic cache is a $\langle$key, value$\rangle$ memory buffer where keys are \textit{embeddings} and values are output lengths.
% Embeddings are generated by embedding models that map text inputs into a low-dimensional continuous vector space and are stored in a vector database~\cite{bang2023gptcache}.
% For each input, the most similar cached input is retrieved using a similarity evaluation function to determine if the cached input matches the input query semantically.

\noindent
\textbf{Other opportunities for using Proxy Models.}
With light proxy models predicting the output token length of incoming LLM serving requests, there can be several potential use cases.
First, the model serving system can dynamically allocate the appropriate amount of memory required for the output, reducing the under-utilization of over-allocation.
The predicted output length can also help in optimizing the buffer size for storing the generated output, minimizing the need for reallocations or unnecessary memory usage.
Second, the predicted output length can inform caching strategies for the generated outputs. Shorter outputs could be cached more aggressively, while longer outputs might be cached selectively or evicted earlier to conserve cache space.
% Cache Partitioning: The cache can be partitioned or organized based on predicted output lengths, allowing for more efficient cache management and eviction policies tailored to different output size ranges.
Third, requests with longer predicted output lengths could be routed to more powerful or dedicated servers or clusters, while shorter requests can be handled by less resource-intensive instances.
Fourth, output length predictions could help with memory offloading from GPU memory to CPU memory or disk, based on the key-value cache access patterns.
However, mispredictions need to be addressed carefully.

\noindent
\textbf{Integration with Speculative Decoding Techniques.}
Speculative decoding or look-ahead decoding accelerates LLM token generation with smaller approximation models~\cite{leviathan2023fast,miao2023specinfer}, multiple decoding heads~\cite{cai2023medusa}, or n-gram generation~\cite{fu2023lookahead}.
% Medusa adds extra heads to LLMs to predict multiple future tokens simultaneously. When augmenting a model with Medusa, the original model stays untouched, and only the new heads are fine-tuned during training.
We did not consider speculative LLM inference because of its extra computational overhead of token generation and verification.

\noindent
\textbf{Integration with vLLM.}
Implementation on vLLM~\cite{vllm} (starting from \texttt{v0.6.2}) the can leverage the priority scheduler by using predicted sequence lengths as priority labels for each incoming request.
This integration allows for more intelligent request prioritization, enabling the system to dynamically adjust scheduling based on anticipated computational requirements (sequence length).

\section{Conclusion}

We have presented \xxx, an efficient model serving scheduler for interactive LLM applications.
The key idea of \xxx is to leverage a proxy-model-based sequence length predictor.
Evaluations on real-world LLM datasets and production workload traces show that \xxx can improve LLM serving JCT by 30.5--39.6\% and throughput by 2.2--3.6\texttimes{} at either no batching, dynamic batching, or continuous batching settings.
\xxx is open-sourced at \href{https://github.com/James-QiuHaoran/LLM-serving-with-proxy-models}{\url{https://github.com/James-QiuHaoran/LLM-serving-with-proxy-models}}.

\section*{Acknowledgments}
We thank the anonymous reviewers for providing their valuable feedback.
This work is supported by National Science Foundation (NSF) under grant No. CCF 20-29049 and by the IBM-ILLINOIS Discovery Accelerator Institute (IIDAI).
Any opinions, findings, conclusions, or recommendations expressed in this material are those of the authors and do not necessarily reflect the views of the NSF or IBM.

\bibliographystyle{plain}
\bibliography{acmart}

\end{document}